%% file: MRP-experience-report.tex
\documentclass[10pt, conference, compsocconf, letterpaper]{IEEEtranv17}
\usepackage{soul}

\usepackage{graphicx}
\usepackage{amsmath}
\usepackage{amssymb}

\usepackage{hyperref}

\usepackage[colorinlistoftodos, textwidth=1cm, shadow]{todonotes}

\begin{document}

%\permission{Permission to make digital or hard copies of all or part of this work for personal or classroom use is granted without fee provided that copies are not made or distributed for profit or commercial advantage and that copies bear this notice and the full citation on the first page. Copyrights for components of this work owned by others than ACM must be honored. Abstracting with credit is permitted. To copy otherwise, or republish, to post on servers or to redistribute to lists, requires prior specific permission and/or a fee. Request permissions from Permissions@acm.org.}
%\conferenceinfo{Middleware '14,}{December 08 - 12 2014, Bordeaux, France}
%\CopyrightYear{2014}
%\copyrightetc{Copyright 2014 ACM \the\acmcopyr}
%\crdata{978-1-4503-2785-5/14/12\ldots\$15.00.\\
%http://dx.doi.org/10.1145/2578726.2578744}

\newcommand{\myparagraph}[1]{\vspace{6pt}\noindent\textbf{#1~}}

\newcommand{\mv}[1]{\ensuremath{\operatorname{\mathit{#1}}}}
\definecolor{dark}{gray}{.6}
\newcommand{\bc}[1]{\textcolor{dark}{#1}}
\newcommand{\hlc}[2][yellow]{ {\sethlcolor{#1} \hl{#2}} } % \hl with a color option
\newtheorem{lems}{Lemma}
\newtheorem{props}{Proposition}
\newtheorem{thms}{Theorem}
\newtheorem{defs}{Definition}
\newtheorem{obs}{Observation}

\title {Stretching Multi-Ring Paxos\\[-1.5ex]}
%\title{Experience Report: 10g Multi-Ring Paxos}
% \author {
%     \IEEEauthorblockN {Submitted as \underline{Experience Report (page limit: 10 pages)}; authors' names omitted for double-blind review.\\
%     The material in this paper has been cleared through the author affiliations.
%     }
%     \\[-3.0ex]
% }

%\thanks{This work was supported in part by the Swiss National Science Foundation under grant number 146714.}

\author {
   \IEEEauthorblockN {
       Samuel Benz%\IEEEauthorrefmark{1}
       \hspace{10mm}
       Leandro Pacheco de Sousa%\IEEEauthorrefmark{1}
       \hspace{10mm}
       Fernando Pedone%\IEEEauthorrefmark{1}
   }
   \IEEEauthorblockA {University of Lugano (USI), Switzerland}
%    \IEEEauthorblockA {\IEEEauthorrefmark{1}University of Lugano (USI), Switzerland}
 %  \\[-3.0ex]
}

\maketitle

\begin{abstract}

Internet-scale services rely on data partitioning and replication to provide scalable performance and high availability.
Moreover, to reduce user-perceived response times and tolerate disasters (i.e., the failure of a whole datacenter), services are increasingly becoming geographically distributed.
Data partitioning and replication, combined with local and geographical distribution, introduce daunting challenges, including the need to carefully order requests among replicas and partitions.
One way to tackle this problem is to use group communication primitives that encapsulate order requirements.
This paper presents a detailed performance evaluation of Multi-Ring Paxos, a scalable group communication primitive.
We focus our analysis on ``extreme conditions" with deployments including high-end 10 Gbps networks, a large number of combined rings (i.e., independent Paxos instances), a large number of replicas in a ring, and a global deployment.
We also report on the performance of recovery under peak load and present two novel extensions to boost Multi-Ring Paxos's performance.

\end{abstract}


%A category including the fourth, optional field follows...
%\category{D.2.8}{Software Engineering}{Metrics}[complexity measures, performance measures]

%\category{C.4}{Performance of Systems}{Fault tolerance}[]
%\category{D.4.7}{Operating Systems}{Organization and Design}[Distributed Systems]

%\terms{Performance}
%\terms{Reliability}

\input{intro.tex}
\input{mrp.tex}
\input{experiments.tex}
\input{related.tex}
\input{conclusion.tex}

\bibliographystyle{ieeetr}
\bibliography{main-2}

\end{document}

%% file: intro.tex
%!TEX root =  MRP-experience-report.tex
\section{Introduction}

%Internet-scale services have become widespread.
Internet-scale services are widely deployed today.
These systems must deal with a virtually unlimited user base, scale with high and often fast demand of resources, and be always available. 
In addition to these challenges, many current services have become geographically distributed.
Geographical distribution helps reduce user-perceived response times and increase availability in the presence of node failures and datacenter disasters (i.e., the failure of an entire datacenter).
In these systems, data \emph{partitioning} (also known as sharding) and \emph{replication} play key roles.

Data partitioning and replication can lead to highly scalable and available systems, however, they introduce daunting challenges.
Handling partitioned and replicated data has created a dichotomy in the design space of large-scale distributed systems.
One approach, known as \emph{weak consistency}, makes the effects of data partitioning and replication visible to the application.
%
%% Weak consistency leads to ``simpler systems", less exposed to impossibility results~\cite{FLP85,GILY2002}, at the expense of more complex and less intuitive applications.
Weak consistency provides more relaxed guarantees and make systems less exposed to impossibility results~\cite{FLP85,GILY2002}.
The tradeoff is that weak consistency generally leads to more complex and less intuitive applications.
The other approach, known as \emph{strong consistency}, hides data partitioning and replication from the application, simplifying application development.
Strong consistency requires ordering requests across the system in order to provide applications with the illusion that state is neither partitioned nor replicated.
%% Weak consistency leads to ``simpler systems", less exposed to impossibility results~\cite{FLP85,GILY2002}, at the expense of more complex and less intuitive applications.
%% Although weak consistency is a natural way to manage the complexity of building scalable and highly available systems, it places the burden on the application designers and users.
%% The other approach, known as \emph{strong consistency}, consists in hiding data partitioning and replication from the application.
%% Strong consistency requires ordering requests across the system in order to provide applications with the illusion that state is neither partitioned nor replicated.

%\pagebreak
Reliably delivering and ordering requests in a distributed system has been extensively studied in the context of group communication (e.g., \cite{BJ87b,HT93}).
Atomic broadcast and atomic multicast, for example, encapsulate the notions of totally and partially ordering requests in a distributed system, respectively.
Among the many group communication protocols proposed in the literature~\cite{DUS04}, this paper focuses on Multi-Ring Paxos~\cite{MPP2012,BMP14}, an atomic multicast protocol based on Paxos ~\cite{lamport1998part}.
Multi-Ring Paxos was designed to scale throughput with the addition of resources, a characteristic uncommon to group communication systems.
Existing atomic broadcast and multicast implementations are typically bounded by the capacity of the nodes that take part in the ordering protocol.
Increasing the number of nodes improves availability, but not performance.
Multi-Ring Paxos scales throughput by composing multiple independent instances of Paxos (i.e., \emph{rings}).
Distributing the load among independent Paxos instances is important to cope with CPU bottlenecks (e.g., as typically happens with the coordinator in Paxos~\cite{RP2010}) and I/O bottlenecks (e.g., acceptor's disks~\cite{MPP2012}).

Multi-Ring Paxos has been shown to perform well in locally and geographically distributed settings~\cite{MPP2012,BMP14}.
In this study, we set out to assess its performance under extreme conditions.
In addition to deepening our understanding about Multi-Ring Paxos, the study also resulted in a number of performance optimizations, which we describe in the following sections.
Our performance assessment was guided by our desire to answer the following questions.
%We were motivated to answer the following questions:
\begin{itemize}
\item Can Multi-Ring Paxos deliver performance that matches high-end networks (i.e., 10 Gbps)?
\item How does a recovering replica impact the performance of operational replicas computing at peak load?
%\item Can Multi-Ring Paxos's skip mechanism, designed to address unbalanced load, handle highly skewed traffic?
\item Multi-Ring Paxos ensures high performance despite unbalanced load in combined rings with a skip mechanism. Can Multi-Ring Paxos's skip mechanism handle highly skewed traffic?
\item How many combined rings in a learner and learners in a ring are ``too many"?
\item Can Multi-Ring Paxos deliver usable performance when deployed around the globe and subject to disasters?
\end{itemize}

This paper makes the following contributions.
First, we review Multi-Ring Paxos's design and introduce two novel techniques, \emph{latency compensation} and \emph{non-disruptive recovery}, which improve Multi-Ring Paxos's performance under strenuous conditions.
Second, we detail Multi-Ring Paxos implementation, evaluate its performance experimentally and answer all questions raised above.
Third, we discuss the lessons we learned during the implementation and evaluation of Multi-Ring Paxos. 

The rest of the paper is structured as follows.
Section~\ref{sec:mrp} presents the design of Multi-Ring Paxos and two techniques that improve its performance.
Section~\ref{sec:experiments} evaluates the performance of Multi-Ring Paxos.
Section~\ref{sec:rwork} reviews related work.
Section~\ref{sec:lessons} discusses our experiences with Multi-Ring Paxos
and Section~\ref{sec:conclusion} concludes the paper.

%We first share our experiences with high throughput, low latency Java applications, and then consider issues more specific to Multi-Ring Paxos.

%Our work makes the following contributions. We demonstrate how 
%Multi-Ring Paxos can scale under extreme conditions; locally up to the network
%speed of 10 Gbps and globally in a world wide distributed environment. Further, 
%we explain and demonstrate how recovery under such load is still possible
%and what kind of tradeoffs can be used. Finally, we evaluate on a local cluster
%how Multi-Ring Paxos scales in the number of subscribed groups and in the number 
%of learners.

%% file: mrp.tex
%!TEX root =  MRP-experience-report.tex
\section{Multi-Ring Paxos}
\label{sec:mrp}

In this section we introduce assumptions and definitions used in Multi-Ring Paxos (Section~\ref{sec:model}), describe the protocol in detail (Section~\ref{sec:design}), present two novel optimizations to Multi-Ring Paxos (Sections~\ref{sec:latency} and~\ref{sec:recovery}), and discuss the protocol's implementation (Section~\ref{sec:impl}).

\subsection{System model and definitions}
\label{sec:model}

We assume a distributed system composed of interconnected processes that communicate through message passing.
Processes may fail by crashing and subsequently recover, but do not experience arbitrary behavior (i.e., no Byzantine failures). 
Processes are either \emph{correct} or \emph{faulty}. 
A correct process is eventually operational ``forever" and can reliably exchange messages with other correct processes. 
In practice, ``forever" means long enough for processes to make some progress (e.g., terminate one instance of consensus).

Multi-Ring Paxos, like Paxos~\cite{Lam01}, ensures safety under both asynchronous and synchronous execution periods. 
To ensure liveness, we assume the system is \emph{partially synchronous}~\cite{DLS88}: it is initially asynchronous and eventually becomes synchronous. 
The time when the system becomes synchronous, called the \emph{Global Stabilization Time (GST)}~\cite{DLS88}, is unknown to the processes. 
Before GST, there are no bounds on the time it takes for messages to be transmitted and actions to be executed. 
After GST, such bounds exist but are unknown. 

Atomic multicast is a communication abstraction defined by the primitives \emph{multicast}$(\gamma,m)$ and \emph{deliver}$(m)$, where $m$ is a message and $\gamma$ is a multicast group.
Processes choose from which multicast groups they wish to deliver messages.
If process $p$ chooses to deliver messages multicast to group $\gamma$, we say that $p$ \emph{subscribes to} group $\gamma$.
Let relation $<$ be defined such that $m < m'$ iff there is a process that delivers $m$ before $m'$.
Atomic multicast ensures that 
(i)~if a process delivers $m$, then all correct processes that subscribe to 
$\gamma$ deliver $m$ \emph{(agreement)}; 
(ii)~if a correct process $p$ multicasts $m$ to $\gamma$ then all correct 
processes that subscribe to $\gamma$ deliver $m$ \emph{(validity)}; and
(iii)~relation $<$ is acyclic \emph{(order)}.
%The order property implies that if processes $p$ and $q$ deliver messages $m$ and $m'$, then they deliver them in the same order.
Atomic broadcast is a special case of atomic multicast where there is a single group to which all processes subscribe.

\subsection{Protocol design}
\label{sec:design}

\begin{figure*}[ht]
 \centering
 \includegraphics[scale=0.8]{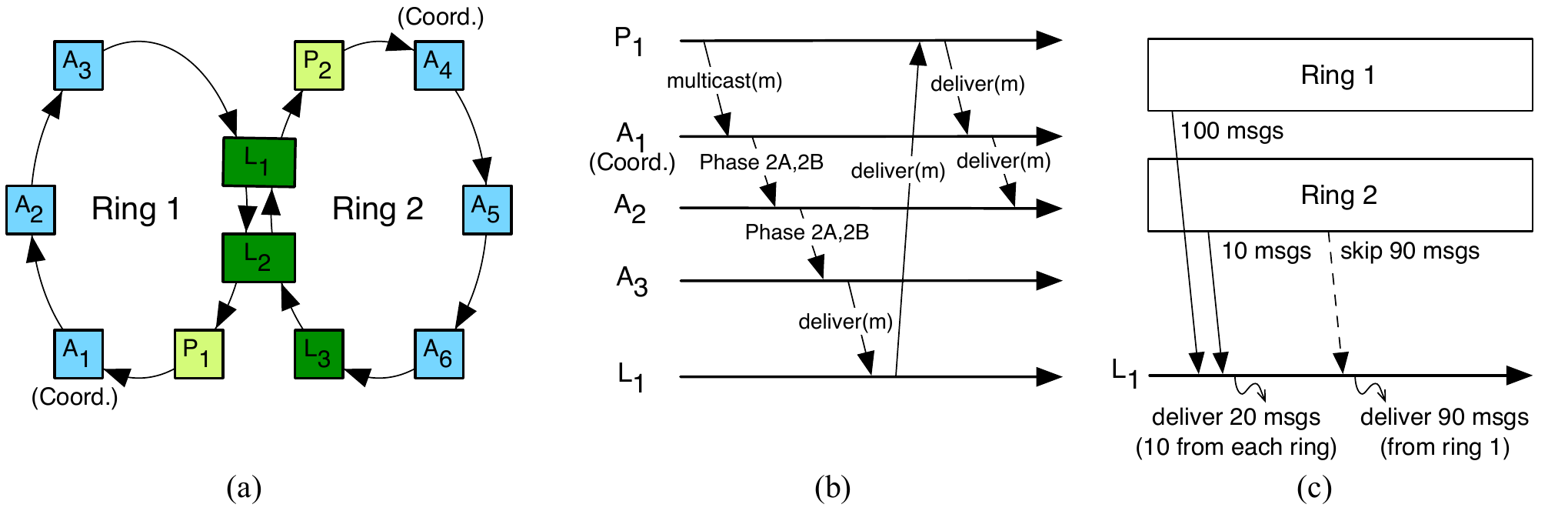}
 \caption{(a)~The various process roles in Ring Paxos disposed in two rings (learners $L_1$ and $L_2$ deliver messages from Rings 1 and 2, and leaner $L_3$ delivers messages from Ring 2 only); (b)~an execution of a single instance of Ring Paxos; and (c)~Multi-Ring Paxos's skip mechanism.}
 \label{fig:rpaxos}
\end{figure*}

Multi-Ring Paxos uses independent Ring Paxos~\cite{RP2010} instances to implement atomic multicast, where each Ring Paxos instance corresponds to an atomic multicast group, as defined in the previous section (see Figure~\ref{fig:rpaxos}).
Ring Paxos efficiently implements the Paxos algorithm~\cite{Lam01} by disposing processes in a ring overlay---for this reason, we sometimes refer to a Ring Paxos instance as a \emph{ring}.
Multi-Ring Paxos ensures ordered delivery of messages using a deterministic round-robin mechanism to merge messages ordered by different rings into a single stream of messages~\cite{MPP2012}.

Multi-Ring Paxos's deterministic merge mechanism ensures that any two messages delivered by two or more processes are delivered in the same order.
If no additional precautions are observed, however, a process will deliver messages at the pace of the slowest multicast group it subscribes to.
To handle unbalanced traffic among rings, a slow ring can skip a configurable number of messages to keep up with the pace of faster rings.
%constant. This approach is easier to implement and to 
%configure, than handling unbalanced traffic in the merge function itself.
The number of ``skip messages" in a ring is determined based on a virtual maximum throughput $\lambda$ that the fastest ring can achieve, measured in messages per second. 
Periodically, each ring coordinator, co-located with its Paxos leader, calculates the number of required skip messages to reach $\lambda$ since the last calculation and multicasts these messages in its ring.
Upon delivering a skip message, a process simply discards it.
Multiple skip messages are grouped in a single Paxos round, containing the number of messages to be skipped.

%Multi-Ring Paxos delivers messages from different groups round-robin. This deterministic
%merge function is simple to implement and shows good performance in their execution. 
%However, this simple approach alone does not compensate different message throughput of
%different groups (rings). The slowest ring determine the overall throughput.

%To handle such a in-balance of a ring throughput, every ring issues ''noop`` commands 
%(skips) when no application traffic is available. The amount of skip messages is
%calculated based on a virtual maximum throughput $\lambda$. Every coordinator, co-located
%with the Paxos leader, determine the amount of required skip messages per time 
%interval $\Delta_t$.

The number of messages to be skipped in a ring at time $t_{\mathit{now}}$, denoted $skips(t_{\mathit{now}})$, is calculated using a referential time $t_{\mathit{ref}}$ and the total number of messages already skipped in the ring, denoted $skipped$, as shown next.
%
%Since all rings delivers the same virtual throughput $\lambda$ per time interval, 
%a delay during the start-up phase between different coordinators, will directly result
%in a wait time (latency) for application messages. Therefor, the skip calculation
%should be based on a previously agreed start time and calculated using local wall clock:
%
\begin{equation}
 skips(t_{\mathit{now}}) = \lambda * (t_{\mathit{now}} - t_{\mathit{ref}}) - skipped
 \label{e:skip}
\end{equation}

Time $t_{\mathit{ref}}$ can be set to 0 or the system's start up time, but it must be the same for the coordinators of all rings.
In order to even out the throughput of the various rings, so that the merge mechanism is effective, every coordinator should periodically recompute equation (1).
%In order to level up the throughput of the various rings, ideally every coordinator should periodically recompute equation~(\ref{e:skip}) at the same time.
In our experiments, we synchronize the coordinator's clocks using a simple NTP service, so that coordinators choose approximately the same $t_{\mathit{now}}$ when recomputing equation~(\ref{e:skip}). 
Note that clock accuracy does not affect the correctness of Multi-Ring Paxos, but has an impact on its performance.

\subsection{Latency compensation}
\label{sec:latency}

The skip calculation described in the previous section is very effective in networks subject to small latencies (e.g., within a datacenter).
However, with large and disparate latencies (e.g., geographical deployments), a late skip message may delay the delivery of messages at a learner (see Figure~\ref{fig:rpaxos}(c)).
This delay might happen even if the number of skip instances is accurately calculated to account for unbalanced traffic among rings.
%% Although the skip calculation described in the previous section is effective in networks subject to small latencies (e.g., within a datacenter), under larger latencies (e.g., geographical deployments), a slow skip message may delay the delivery of messages at a learner (see Figure~\ref{fig:rpaxos}(c)), even if the number of skip instances is accurately calculated to account for unbalanced traffic among rings.
%
%In \ref{e:skip} every coordinator calculates the amount of skip messages to be sent
%based on $time{now}$. This messages however, are not arriving at the deterministic
%merge function at the same time. In a global environment, the difference between
%the two skip messages from different rings can be 80 ms or even more. The skip 
%messages calculated with \ref{e:skip} will be delivered in that example with the
%exact same time difference of 80 ms. In other words, local commands will be see
%an additional latency of this 80 ms.
%
We overcome this problem by revisiting the skip mechanism to take into consideration the approximate time skip messages need to reach their concerned learners.
In equation~(\ref{e:skip2}), $avg\_delay$ is an approximated average of the delays between the ring coordinator and the ring learners.
The intuition is to skip additional messages to make up for the time it takes for a skip message to arrive at the learners.

\begin{equation}
 skips(t_{\mathit{now}}) = \lambda * (t_{\mathit{now}} - t_{\mathit{ref}} - avg\_delay) - skipped
 \label{e:skip2}
\end{equation}

%The coordinators in all local rings will compensate the WAN $latency$ (\ref{e:skipnew}).
%This ensures, that the merge function ``sees'' the latency independent amount of
%skips at every time.
%
%The latency measurement can be done at the merge function. By simply including
%a time stamp of the created skip messages, every merge function (learner) has a
%good estimate of this latency. This estimates can be sent to the local coordinator.
%Since we assume that the WAN latency is much higher than the local latency, the 
%different estimates seen by multiple learners in a ring doesn't have a big impact. 

\subsection{Non-disruptive recovery}
\label{sec:recovery}

Recovering a failed learner in Multi-Ring Paxos, as described in \cite{BMP14}, boils down to (a)~retrieving and installing the most recent service's checkpoint and (b)~recovering and executing commands that are not included in the retrieved snapshot, the \emph{log tail}.
While this procedure can be optimized in many ways~\cite{bessani2013efficiency}, recovery in Multi-Ring Paxos is inherently subject to a tradeoff that involves the frequency of checkpoints and the size of the log tail: frequent checkpoints result in smaller log tails and, conversely, infrequent checkpoints lead to larger log tails.

Since checkpoints tend to slow down service execution, reducing the frequency of checkpoints seems desirable.
However, restricting the log tail size is equally important because retrieving commands from the log during recovery has negative effects on the service's performance---this happens because acceptors must participate in new rounds of Paxos and at the same time retrieve values accepted in earlier rounds (i.e., the log tail).
We have experimentally assessed that even under moderate load the recovery traffic drastically affects performance (see Section~\ref{sec:nondisrec}). 

To minimize disruption of service performance during normal service execution and recovery of a learner, we revisited Multi-Ring Paxos's original recovery mechanism~\cite{BMP14}. 
With the new method, a recovering learner starts by caching new ordered messages.
This silent procedure does not place acceptors under additional stress.
The replica then must retrieve a valid checkpointed state from another replica (or from remote storage), that is, a checkpoint that contains all commands that precede the cached commands.
With a valid checkpoint, the replica can apply the cached commands not in the the checkpoint and discard the ones already in the checkpoint.
This procedure prioritizes performance during normal operation but it may increase the time needed to recover a learner.

\subsection{Implementation}
\label{sec:impl}

URingPaxos,\footnote{https://github.com/sambenz/URingPaxos} our Multi-Ring Paxos prototype, is entirely implemented in Java with communication relying on TCP. 
The ring and configuration management are handled by Zookeeper~\cite{hunt2010zookeeper}.
A URingPaxos node can play multiple Paxos roles (e.g., proposer, acceptor, learner). 
Upon starting, the node registers its IP address and its intended roles with Zookeeper. 
The node is then informed by Zookeeper about the endpoints it should connect to to form a ring. 
If a node crashes, Zookeeper will inform all nodes about the topology change.
%The network transport is implemented with TCP.
Acceptors have access to stable storage. 
Depending on the required guaranties, back-ends for in-memory storage and synchronous and asynchronous on-disk storage are available.
The first acceptor in the ring is elected the coordinator. 

We implemented our own internal serialization based on byte buffers and ensure that at most one object is created per received item.
To avoid heavy garbage collection work, the current implementation pre-allocates an array of byte buffers. 
%Replacing the content of this buffers never triggers a major garbage collection effort.
As a result, under normal load, garbage collection does not significantly impact performance. 
%For all experiments, we use Java default settings. 

%%%%%%%%%%%%%%%%%%%%%%%%%%%%%%%%%%%%%%%%%%%%%%%%%%%%%%%%%%%%%%%%%%%%%%%%%%%%%%%%

%% file: experiments.tex
%!TEX root =  MRP-experience-report.tex
\section{Experimental Evaluation}
\label{sec:experiments}

In this section, we explain our goals and methodology, describe our experimental setup, 
detail Multi-Ring Paxos configuration, and present and comment our findings.
%experimentally assess various aspects of the performance of 
%multi-Ring Paxos under extreme conditions.

\subsection{Objectives and methodology}

We aim to assess the behavior of Multi-Ring Paxos under a range of ``extreme" conditions, including wide-area channels and high-performance links.
Since we do not have access to an experimental environment that simultaneously accommodates all these characteristics,
 %(e.g., our wide-area links have limited bandwidth capacity, our 10 Gb network has only a few nodes), 
 we conducted our experiments in different environments and workload settings, as described next.

%\vspace{-4mm}
\begin{itemize}
% \item \hl{Implementation efficiency}
%\begin{itemize}
\item We scale the number of rings to achieve high performance in a high-end 10 Gbps network (Section~\ref{sec:10gbps}).
\item We evaluate the impact of a recovering replica on the performance of operational replicas under peak load (Section~\ref{sec:nondisrec}).
%\item We stress Multi-Ring Paxos skip mechanism with highly skewed traffic (Section~\ref{sec:skipcost}).
%\item We study the impact of many learners in a single ring (Section~\ref{sec:largerings}).
%\item We assess the impact of a global ring and a disaster failure in a geographically distributed deployment (Section~\ref{sec:orderglobe}).
%\end{itemize}
%
% \item \hl{Protocol efficiency}
%\begin{itemize}
%\item We scale the number of rings to achieve high performance in a high-end 10 Gbps network (Section~\ref{sec:10gbps}).
%\item We evaluate the impact of a recovering replica on the performance of operational replicas under peak load (Section~\ref{sec:nondisrec}).
\item We stress Multi-Ring Paxos skip mechanism with highly skewed traffic (Section~\ref{sec:skipcost}).
\item We study the impact of many learners in a single ring (Section~\ref{sec:largerings}).
\item We assess the impact of a global ring and a disaster failure in a geographically distributed deployment (Section~\ref{sec:orderglobe}).
%\end{itemize}
\end{itemize}

%We focus on two message sizes: small messages of 200 bytes and large messages of 32 Kbytes for performance measurements.
%We use small messages to stress garbage collection and CPU, and large messages to determine peak throughput (bits per second).
%
%\hl{Recovery under peak load is evaluated under more practical conditions wit 4 and 8-Kbyte commands.}

%Deployments with a single ring correspond to Ring Paxos, our baseline.

\subsection{Experimental setup}

The local-area network experiments (i.e., within a datacenter) were performed in two environments:
(a)~A cluster of 4 servers equipped with 32-core 2.6 GHz Xeon CPUs and 128 GB of main memory. 
These servers were interconnected through a 48-port 10-Gbps switch with round trip time of 0.1 millisecond. 
(b)~A cluster of 24 Dell PowerEdge 1435 servers and 40 HP SE1102 servers connected through two HP ProCurve 2910 switches with 1-Gbps interfaces.
The globally distributed experiments (i.e., across datacenters) were performed on Amazon EC2 with instances in 5 different regions.
We used r3.large spot-instances, with 2 vCPU and 15 GB DRAM.
To avoid disk bottlenecks, all experiments were executed with in-memory storage. 
A detailed evaluation of Multi-Ring Paxos under different storage conditions
can be found in \cite{BMP14}.

\subsection{Multi-Ring Paxos configuration}

Multi-Ring Paxos has three configuration parameters~\cite{MPP2012}: $M$, $\lambda$ and $\Delta_t$.
$M$ is the number of messages delivered (or skipped) contiguously from the same single ring; if not stated otherwise, we use $M=1$.

We have empirically determined that $\lambda$, the virtually maximum throughput of a ring, should be set a bit higher than the actual maximum achievable performance.
Too high $\lambda$ values lead to wasted CPU cycles in the deterministic merge function; too low $\lambda$ values cap performance.

Parameter $\Delta_t$ determines how often skip messages are proposed in a Paxos instance.
In general, small values for $\Delta_t$ are preferred, to reduce the latency of 
actual messages; too low $\Delta_t$ values, however, waste Paxos instances and 
introduce additional overhead in the system.

\subsection{Scaling up in a local 10 Gbps network} % local scalability
%\subsection{Scale-up Throughput} % local scalability
\label{sec:10gbps}

In this section, we evaluate the scalability of Multi-Ring Paxos in a local 10 Gbps network environment.

\textbf{Setup.} 
We perform two sets of experiments, one with 200-byte messages and another with 32-Kbyte messages.
For each message size, we increase the number of rings from 1 (i.e., Ring Paxos) up to 10.
%
%In this experiment, we increase the number of rings at a single learner while sending 200-byte or 32-Kbyte messages to all rings. 
Four servers are involved: one server runs one proposer and one acceptor per ring, two other servers play the role of acceptors only, with one acceptor deployed per ring; the last server runs a learner, which subscribes to up to 10 rings.
The proposer in each ring uses multiple threads (20), one thread per client.
%The proposer is multithreaded (20 threads), with one thread per clients.
%The proposer in each ring is multithreaded (with 20 threads). 
%Small messages are batched together up to 32 Kbytes.
We report peak throughput, measured at the learner.

\begin{figure*}[t]
  \begin{center}
    \begin{tabular}{c}  
      \includegraphics[width=\textwidth]{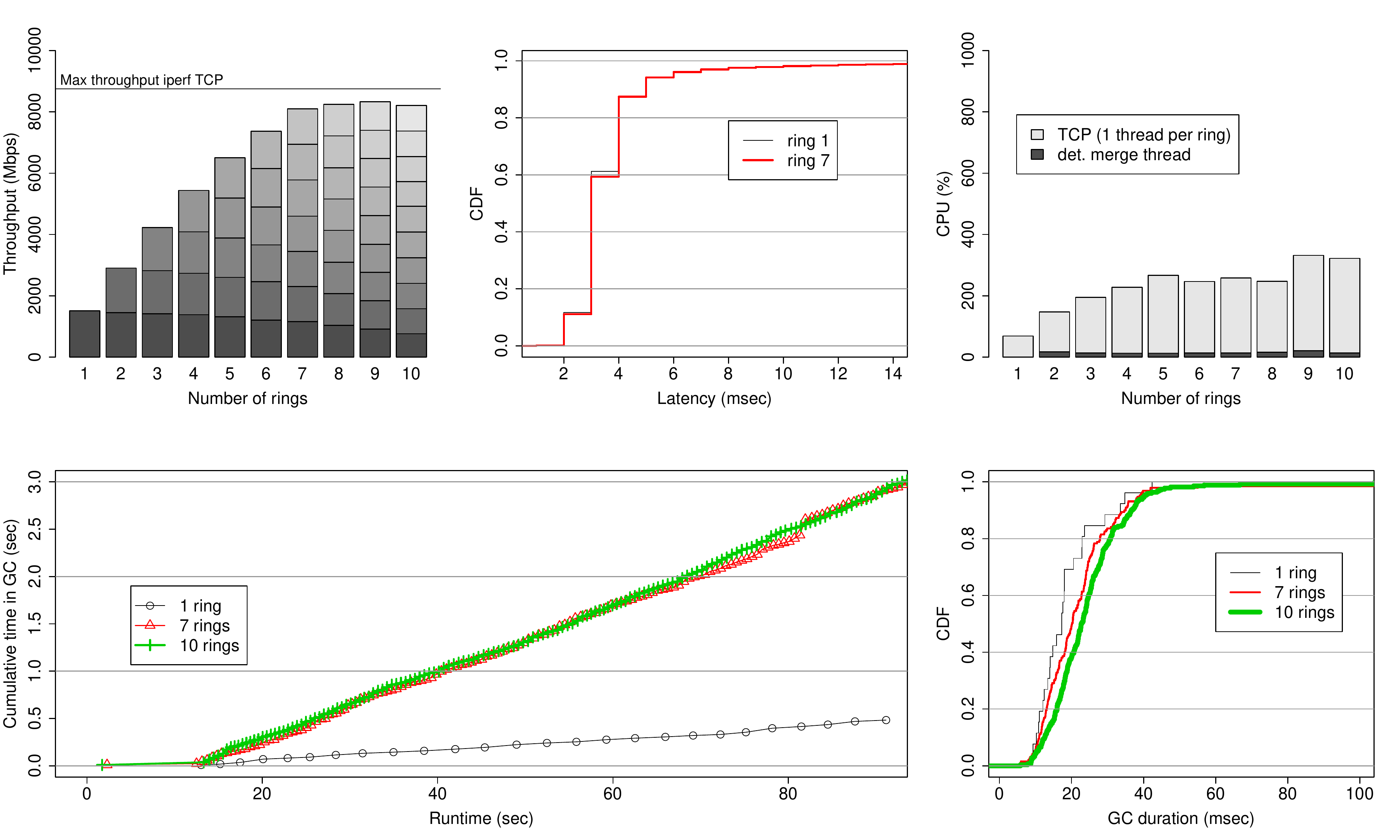} 
    \end{tabular}
\caption{Scaling up Multi-Ring Paxos in a 10 Gbps network. The graphs show the aggregate and per ring throughput in megabits per second for 32-Kbyte messages (top left); the latency CDF, measured in 1-millisecond buckets (top center); the CPU usage (top right); garbage collection activity during some executions (bottom left) and the CDF of the duration of the Java garbage collection work (bottom right). All measurements performed at the learner process.}
%\vspace{-6mm}
    \label{fig:10g-mrp}
  \end{center}
\end{figure*}

\begin{figure*}[t]
  \begin{center}
    \begin{tabular}{c}  
      \includegraphics[width=\textwidth]{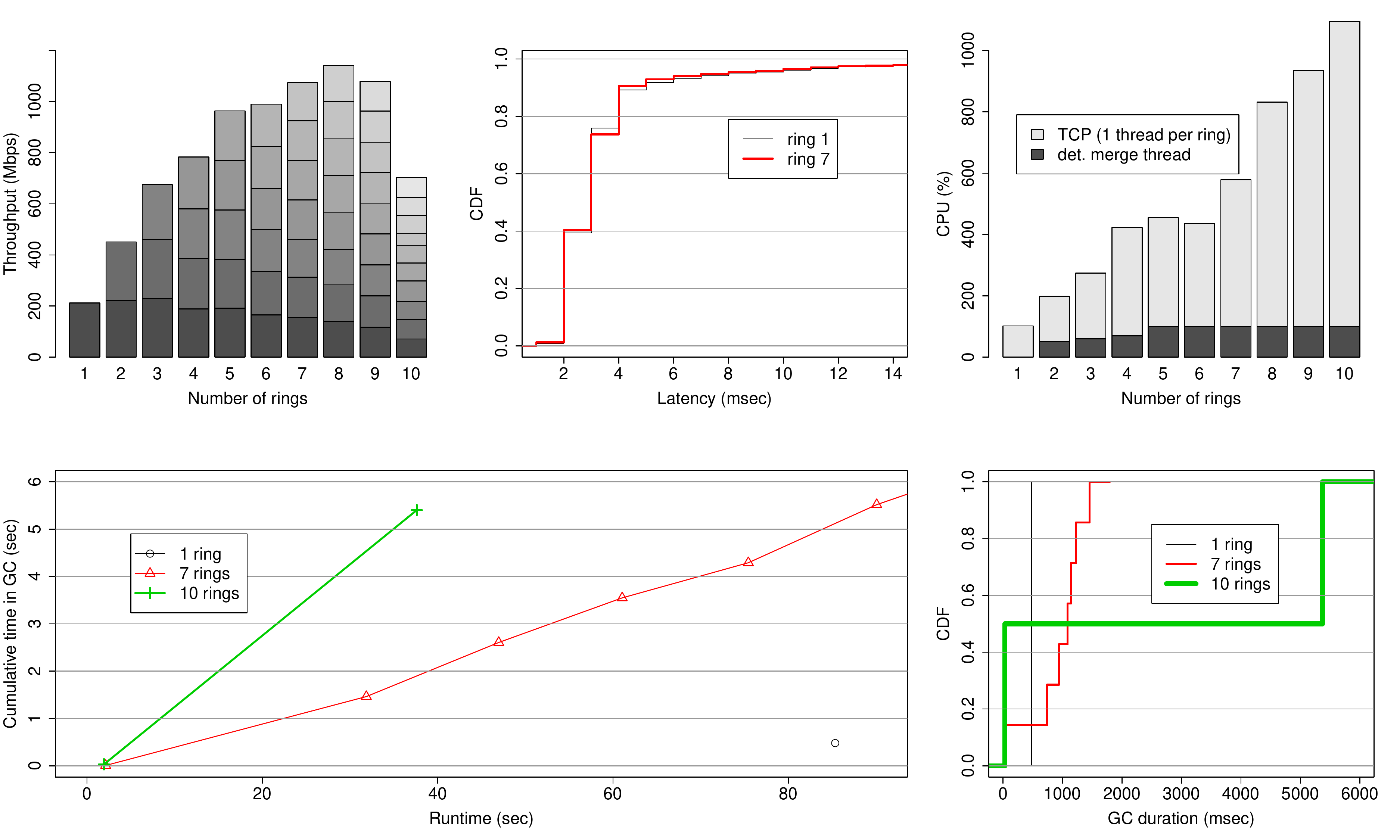} 
    \end{tabular}
\caption{Scaling up Multi-Ring Paxos in a 10 Gbps network. The graphs show the aggregate and per ring throughput in megabits per second for 200-byte messages (top left); the latency CDF, measured in 1-millisecond buckets (top center); the CPU usage (top right); garbage collection activity during some executions (bottom left) and the CDF of the duration of the Java garbage collection work (bottom right). All measurements performed at the learner process.}
%\vspace{-6mm}
    \label{fig:10g-200-mrp}
  \end{center}
\end{figure*}

\textbf{Results.} 
Figures~\ref{fig:10g-mrp} and~\ref{fig:10g-200-mrp} (top left graphs) show, respectively, that Multi-Ring Paxos reaches peak performance with 9 rings for large messages and with 8 rings for small messages.
With large messages, Multi-Ring Paxos reaches 8.41 Gbps, very close to 8.75 Gbps, the maximum usable TCP throughput (i.e., without TCP/IP headers) we could produce with \textit{iperf}.\footnote{http://iperf.sourceforge.net/}
With small messages, Multi-Ring Paxos achieves about 570 K messages per second.
We also report the latency CDF, measured in 1-millisecond buckets, for the peak load (top center graphs) and the CPU consumption at the learner (top right graphs).
The 90-th latency percentile under these conditions is below 5 milliseconds. 
The protocol is network-bound with large messages and CPU-bound with small messages.
(Since there is one communication thread per ring at the learner, 10 rings can use up to 1000\% CPU.)

In both experiments we can see (top left graphs) that as the number of rings a learner subscribes to increases, the throughput achieved by each ring decreases. 
This happens because the load in the learner's Java virtual machine increases with each new ring, slowing down the learner. 
In Multi-Ring Paxos, a slow process reduces the overall traffic, as a result of flow control.
This effect can be seen in the garbage collection logs (bottom two graphs) of the two runs.

\subsection{Non-disruptive recovery under peak load}
\label{sec:nondisrec}

To evaluate our optimized recovery procedure, we deploy a simple key-value store service, implemented on top of Multi-Ring Paxos~\cite{BMP14}.
Our key-value store service implements commands to insert and remove tuples of arbitrary size, read and update an existing entry, and query a range of tuples.
Replicas use a copy-on-write data structure to allow checkpoints in parallel with the execution of commands. 
%As workload, we use update commands of 1024 bytes~\cite{cooper2010benchmarking}.

\textbf{Setup.} 
The experimental setup uses a ring with 3 nodes, each acting as an acceptor and a learner (i.e., replica). 
Four clients (each with 150 threads) submit 1024-byte update requests to the replicas through Multi-Ring Paxos~\cite{cooper2010benchmarking}. 
Each replica executes every request and replies back to the client using UDP. 
Every replica periodically checkpoints its state into a distributed file system,\footnote{http://www.xtreemfs.org/} accessible to all replicas.
The state checkpointed by a replica has 1.5 million entries.

\begin{figure}[t]
  \begin{center}
    \begin{tabular}{c}  
      \includegraphics[width=\columnwidth]{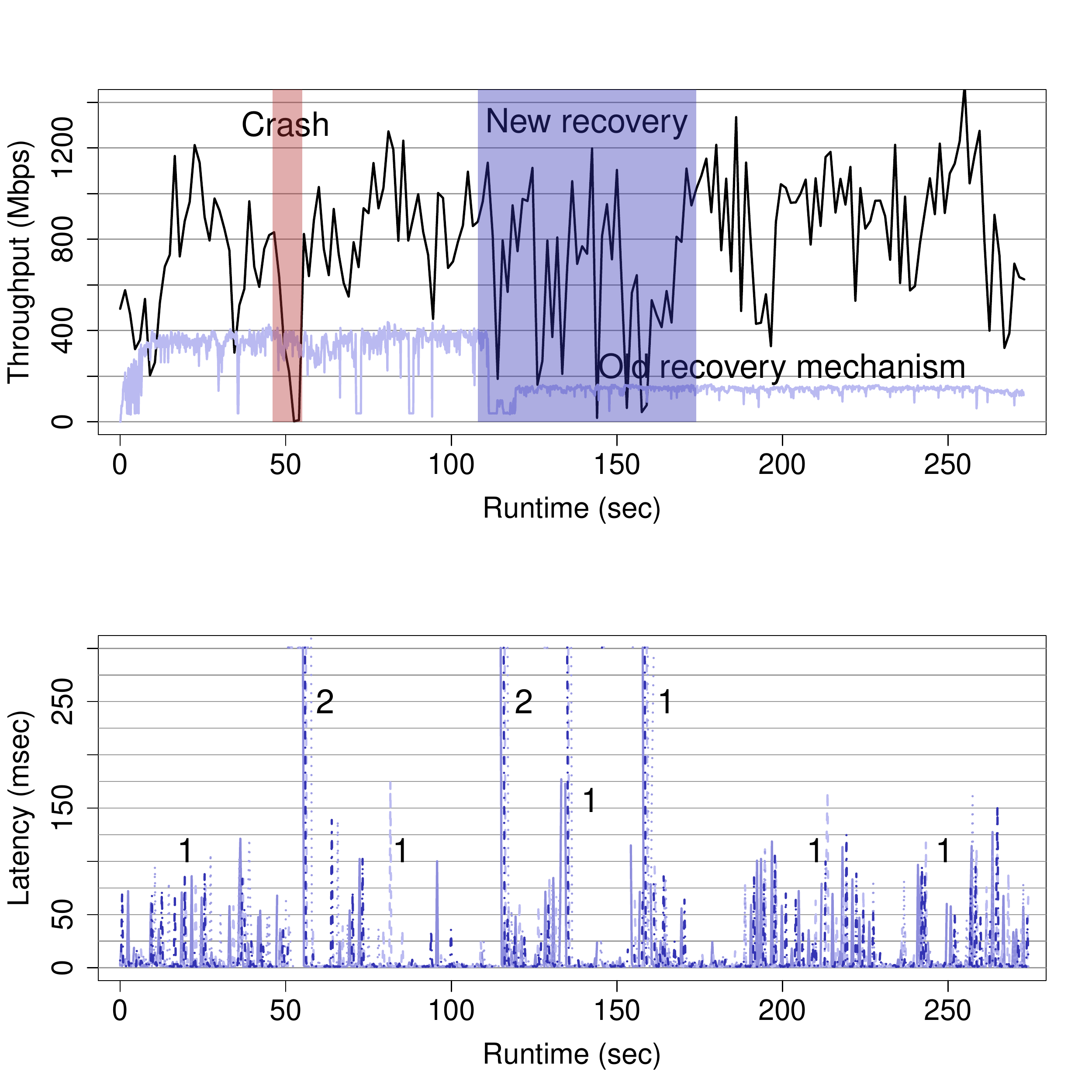} 
    \end{tabular}
\caption{Recovery of a key-value store snapshot with 1.5 million entries. Throughput of Multi-Ring Paxos's new and old recovery protocols (top) and latency of new recovery protocol (bottom, where ``1" identifies garbage collection events and ``2" identifies ring management events).}
%\vspace{-6mm}
    \label{fig:recovery}
  \end{center}
\end{figure}

\textbf{Results.}
Figure~\ref{fig:recovery} shows the behavior of Multi-Ring Paxos's new non-disruptive recovery under maximum load, which for 1024-byte values is around 800 Mbps.
For comparison, we also depict the behavior of the old recovery protocol under lower load, around 400 Mbps, since the old protocol cannot sustain higher load.
Around 45 seconds into the execution, we crash one of the replicas, which starts recovery around time 110.
With the new recovery protocol, the average throughput during recovery is 78\% of the throughput under normal operation.
%
%Recovery does not impact maximal throughput over a long period. The average throughput during recovery is 78\% compared to non recovery time.
Performance troughs are due to garbage collection (events labelled ``1" in the graph) and ring management (event with label ``2").
%All visible effects for a single replica writing parallel snapshots and three replicas of which one is recovering is due to garbage collection (1/3/4).
%In fact, in all cases, the replicas must keep a lot more data in memory than the snapshot size. 
Since processes communicate in a ring, a pause in any of the nodes (e.g., due to garbage collection) can have a visible effect on throughput.
%Since processes communicate in a ring, the pause of any node has an effect on throughput.
%The visible GC impact after 130s is the memory cleanup of the recovered replica (3/4). 
%To not influence the maximum throughput, this replica cached all commands up to a recent snapshot could be installed. 
%After that point, all already applied commands could be freed from memory (4). 
%%A similar effect is visible around 45s and 100s. 
%Under such a load, almost every object is updated during the write process. 
%Freeing this amount of memory after finished checkpointing (1) is visible in the overall throughput.
The fact that the recovering learner has to batch new commands and that replicas have to use multiple (in-memory) copy-on-write data structures forces us to use large heaps, which lead to longer and unpredictable garbage collection pauses (see Section~\ref{sec:lessons}).

%The red line shows the behavior of Multi-Ring Paxos's old recovery protocol under moderate load (1 client).

% \hl{The passive participation (queuing) of the recovering learner and the use of multiple copy-on-write data structure forced us to use very big Java heaps. 
% As bigger the Java heap, as more unpredictable garbage collection pauses we could observe} (Section~\ref{sec:lessons}).

% \begin{figure}[t]
%   \begin{center}
%     \begin{tabular}{c}  
%       \includegraphics[width=\columnwidth]{../experiments/recovery/old-vs-new.pdf} 
%     \end{tabular}
% \caption{Compare classical recovery with the new approach presented in this study under moderate load.}
% %\vspace{-6mm}
%     \label{fig:newold}
%   \end{center}
% \end{figure}

\subsection{The skip mechanism under highly skewed traffic}
%\subsection{Scaling up with many groups}
\label{sec:skipcost}

In Multi-Ring Paxos, learners can subscribe to any combination of existing rings. 
Unbalanced traffic across rings is compensated with the skip mechanism.
%This experiment will assess the behavior of a learner under such a load.
In this experiment, we assess the overhead of the skip mechanism on highly skewed traffic.

\textbf{Setup.} 
This experiment was conducted in a local cluster with a 1 Gbps network.
In this experiment, a single learner subscribes to multiple rings.
Each ring is composed of three acceptors and the learner. 
In order to assess the protocol's inherent latency without any queuing effects, we consider executions with a single client.
We varied the number of rings from 1 up to 32.
Except for the configurations with 16 and 32 rings, we deploy one acceptor per node.
For the experiments with 16 rings, there are two acceptors per node; with 32 rings, there are four acceptors per node.
To assess the efficacy of the skip mechanism, the client submits 200-byte messages to one of the rings; the other rings rely solely on the skip mechanism.
%Since we are interested in highly skewed traffic, we deploy one client only, which submits 200-byte messages to a single ring; all other rings rely on the skip mechanism.
In these experiments, $\lambda$ was set to 5 milliseconds.

\begin{figure}[t]
  \begin{center}
    \begin{tabular}{c}  
      \includegraphics[width=\columnwidth]{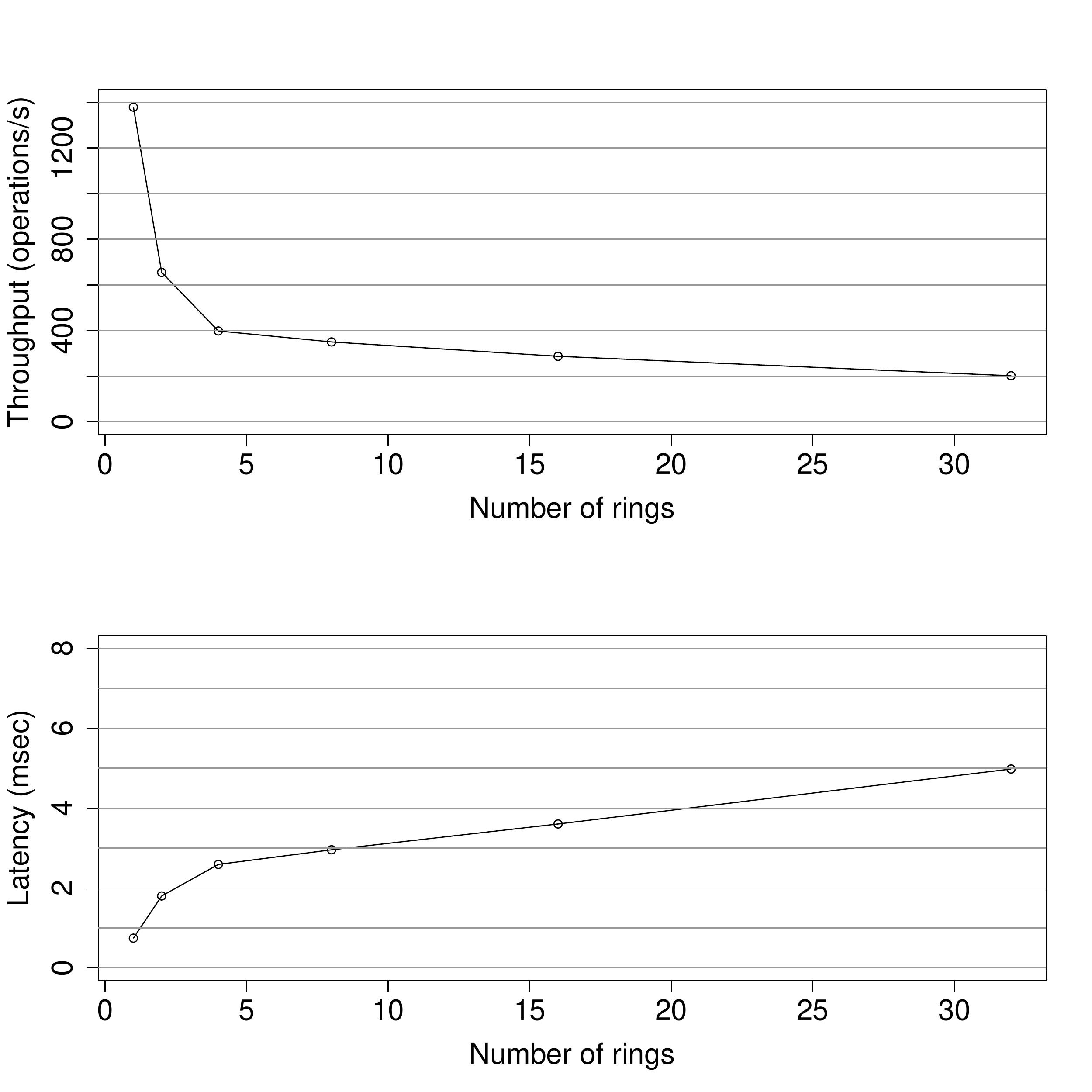} 
    \end{tabular}
\caption{Impact of the number of groups (rings) a learner subscribes to on throughput and latency (since there is a single client, from Little's law throughput is the inverse of latency).}
%\vspace{-6mm}
    \label{fig:rings}
  \end{center}
\end{figure}

\textbf{Results.} 
The most visible impact in Figure~\ref{fig:rings} is the transition from one to two rings. 
One ring is not constrained by any synchronization and can achieve the lowest latency.
Additional rings introduce an overhead, that eventually increases linearly with the number of rings.
Since we have one client only, from Little's law~\cite{Jai91}, the throughput is the inverse of the latency. 
%The latency increases as expected by adding more rings, since the probability that 
%one ring did not yet received a skip message also increases. After 32 ring, we 
%reach the upper bound of the latency = $\lambda = 5 ms$.

\subsection{The performance of large rings}
%\subsection{Scaling up with many learners}
\label{sec:largerings}

Although fault-tolerant deployments usually require a few replicas (e.g., three to five), having a large number of learners inside a single ring is also useful.
One common example is to use a large ring containing all replicas from smaller rings.
In this case, each small ring would encompass some partition of the service's state and the large ring could be used to send commands concerning multiple partitions~\cite{BMP14}.
%\hl{Like in the previous experiment, we are interested in the protocol behavior and not maximum throughput. Therefore we keep the client load small.}
%Messages on the large ring are ordered across all partitions and can be used to implement the classic definition of multicast (sending one message to multiple groups).

%This scheme allows for messages to be destined to any combination of rings while still keeping ordering guarantees.

%we wanted to understand the effects of many learners on the performance of a ring.
% Deploying many learners in a single large ring can be useful when the large ring aggregates all replicas of smaller rings.
% For example, the large ring can be used to propagate meta data to coordinate the activities in the smaller rings, which represented a partitioned state of a service.
%This can be the case of a large ring that aggregates all replicas in smaller rings.
%Sometime a deployment of Multi-Ring Paxos requires a single ring with many learners.
%While normally three to five learners, acting as replicas, are enough, especially 
%in a local partitioned environment a lot more learners per ring are required. E.g. the
%ring which spawns all replicas.

\textbf{Setup.} 
This experiment was conducted in a local cluster with a 1 Gbps network.
There is one ring with three acceptors and an increasing number of learners. 
With up to 32 learners, we deployed each learner in an HP SE1102 server; we deployed additional learners in the weaker Dell PowerEdge servers.
To assess the protocol's latency in the absence of any queuing effects due to contention, we consider executions with a single client, which sends messages to a proposer in the ring.
%The client proposes messages are proposed from a client that is not part of the ring (i.e., the client sends messages to a proposer in the ring). 
%No batching is used.

\begin{figure}[t]
  \begin{center}
    \begin{tabular}{c}  
      \includegraphics[width=\columnwidth]{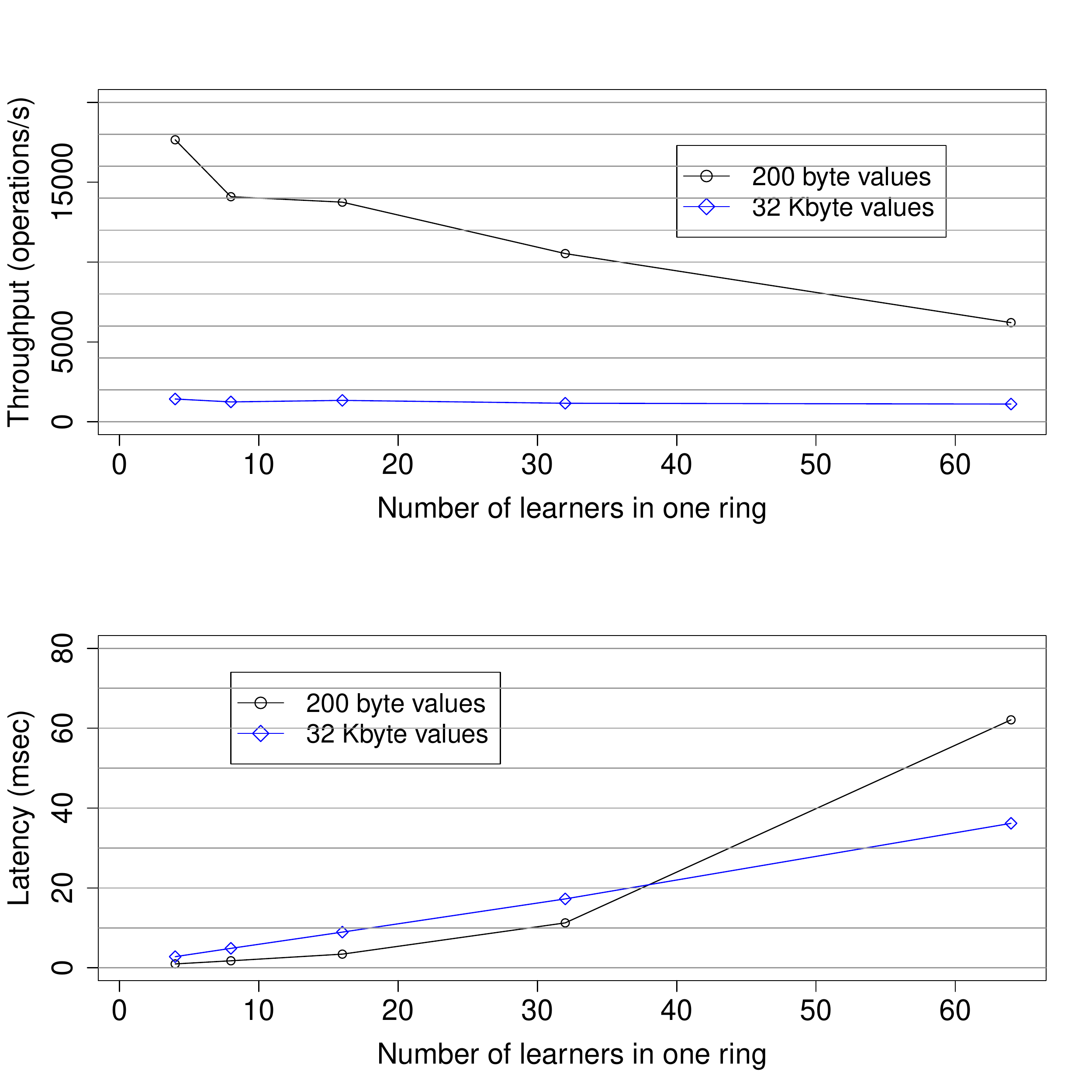} 
    \end{tabular}
\caption{Impact of adding many learners to a single ring. Every node in the ring
adds linearly latency.}
%\vspace{-6mm}
    \label{fig:learners}
  \end{center}
\end{figure}

\textbf{Results.} 
Figure~\ref{fig:learners} shows the effect of adding learners to a single ring.
Like in the previous experiment, the single client results in throughput inversely proportional to latency. 
Further, ring communication in Multi-Ring Paxos linearly adds latency with every additional node.
The sharp bend in latency after adding 32 learners is caused by the weaker nodes, which become CPU-bound with small messages.

\subsection{Ordering messages across the globe} % global scalability
\label{sec:orderglobe}

In this section we evaluate the global scalability and fault tolerance of Multi-Ring Paxos. 
The goal is to show that having a large global ring, which allows to send ordered commands to geographically distributed partitions (local rings), does not slow down local traffic.
We also evaluate the effect of a data center outage during runtime.

% Since we have only the data for 1 run with 5 DC we can not compare the impact of
% adding more and more DC like in the MRP Recovery paper! -> so focusing on DC failure

\textbf{Setup.} 
For this experiment, we used Amazon EC2 instances.
We deployed 5 local rings, each in its own region: us-west-1 (N. California), us-west-2 (Oregon), eu-west-1 (Ireland), ap-southeast-1 (Singapore), ap-southeast-2 (Sydney).
All nodes in each local ring are placed on the same availability zone.
We also deployed a global ring, composed of three acceptors (placed in separate regions) and all learners from each of the local rings.
This deployment allows for progress even in the presence of a disaster taking down an entire datacenter.
We simulated a datacenter outage by forcibly killing all processes belonging to one of the regions containing an acceptor of the global ring.

%% This experiment runs on Amazon EC2. We use 5 regions: us-west-1 (N. California), us-west-2 (Oregon), eu-west-1 (Ireland), ap-southeast-1 (Singapore), ap-southeast-2 (Sydney).
%% Inside of a region, all servers are placed in a single availability zone. 
%% The three acceptors of the global ring are placed in different regions. 
%% This deployment ensures progress despite the disastrous crash of a complete region, which we induce after 30 seconds of run time.

\begin{figure}[t]
  \begin{center}
    \begin{tabular}{c}  
      \includegraphics[width=\columnwidth]{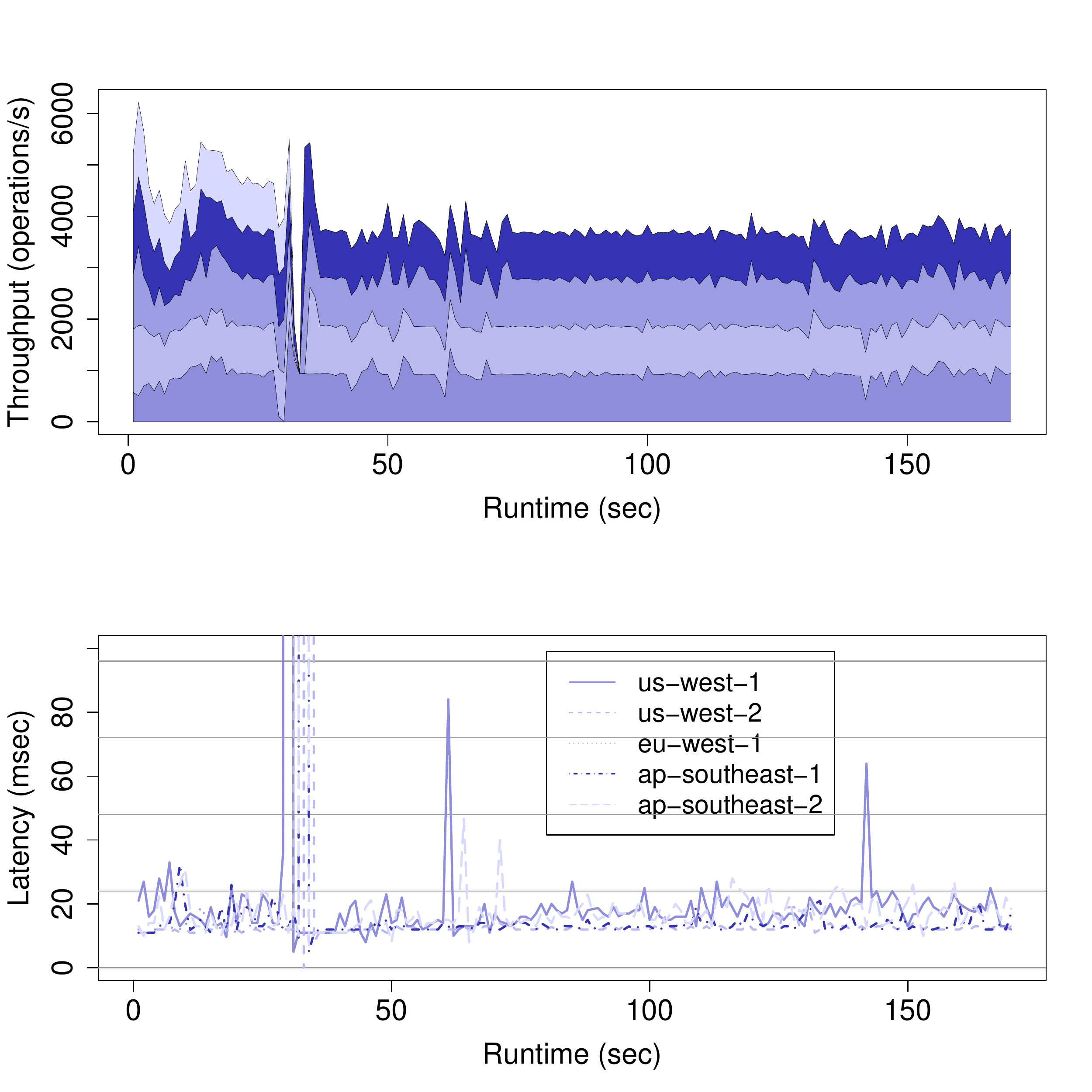} 
    \end{tabular}
\caption{Impact of a data center outage after 25s of execution in the performance of a global Multi-Ring Paxos deployment.}
%\vspace{-6mm}
    \label{fig:ec2}
  \end{center}
\end{figure}

\textbf{Results.} 
We first evaluate the fault tolerance of Multi-Ring Paxos.
Figure~\ref{fig:ec2} shows the throughput in each of the local rings, using messages of 32 Kbytes.
We can see that, despite the outage of a complete region (at around 25 seconds of execution), the remaining rings maintain normal traffic after a short disruption caused by the global ring reconfiguration.

\begin{figure}[t]
  \begin{center}
    \begin{tabular}{c}
      \includegraphics[width=\columnwidth]{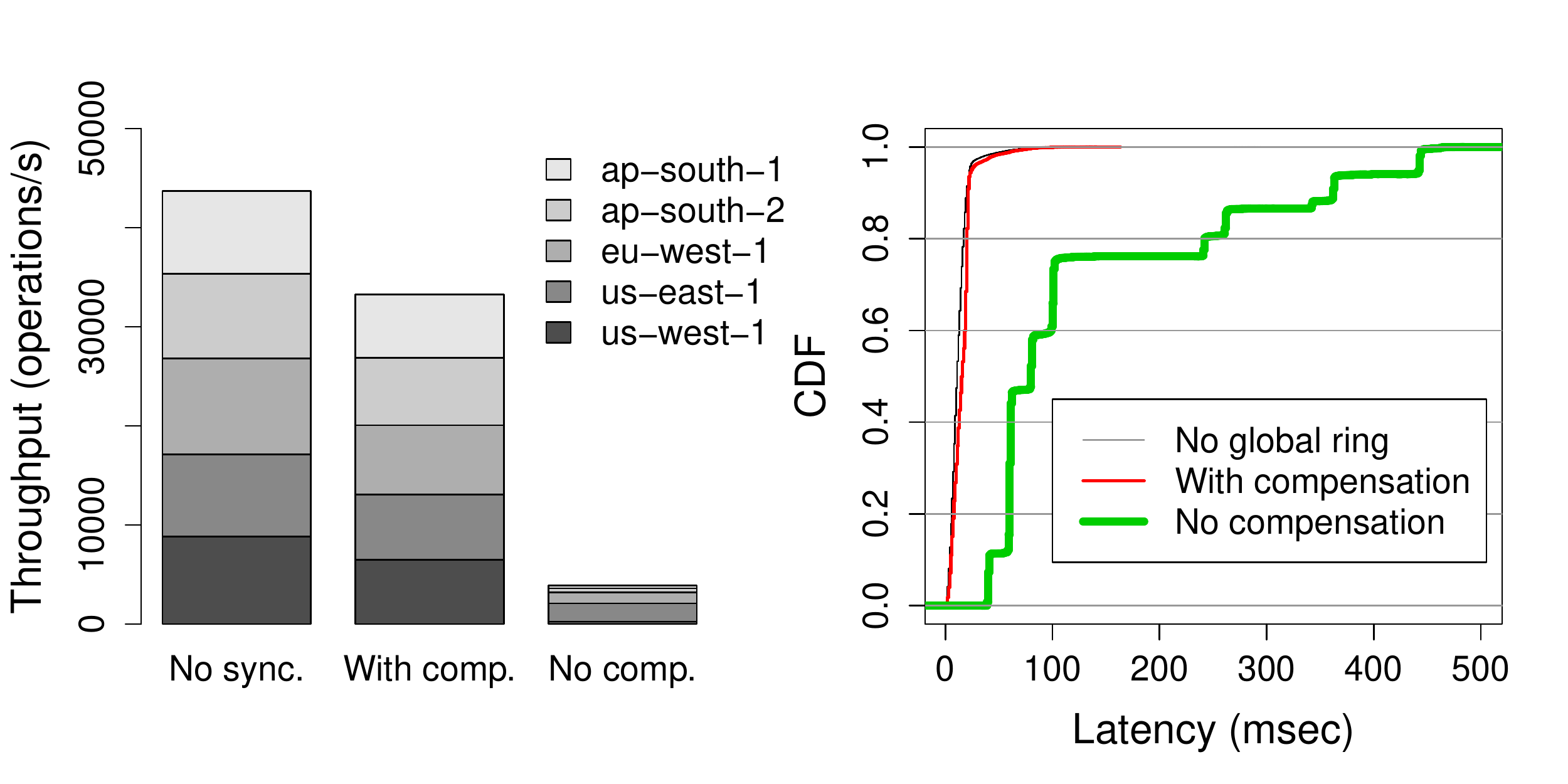} 
    \end{tabular}
\caption{Impact of a global ring to local maximum throughput with and without latency compensated skip calculation.}
%\vspace{-6mm}
    \label{fig:ec2skip}
  \end{center}
\end{figure}

To assess the impact of a global ring on the performance of local rings, we conducted a few other experiments using the same deployment of 5 datacenters, each with a local ring.
We consider a baseline case with local rings only (i.e., no global ring) and setups with a global ring synchronizing all nodes, with and without latency compensation (Section~\ref{sec:latency}).
We use the same load (number of clients) in all three cases, roughly 80\% of the peak throughput for the case with compensation enabled, with 200-byte messages.
Figure~\ref{fig:ec2skip} shows the throughput obtained in each case and the latency CDF.
The local throughput went down by around 23\% with a global ring connecting all the nodes.
The results also show that compensating the latency difference between rings is fundamental.
The ``steps'' visible in the latency CDF for the scenario with no compensation reflect the latency difference across rings.
Since the datacenters also form a (global) ring in the way they are arranged, the ordering and latency between them affects when a message from the global ring is delivered at each datacenter.
Having the compensation mechanism allows for this latency difference to be masked.

%We have also deployed the five local rings independently, without a global ring synchronizing them.
%We observed in Figure~\ref{fig:ec2skip} that having a global ring did not affect local ring throughput and only had a% small effect on latency. 

%%%%%%%%%%%%%%%%%%%%%%%%%%%%%%%%%%%%%%%%%%%%%%%%%%%%%%%%%%%%%%%%%%%%%%%%%%%%%%%%

%% file: related.tex
\section{Related work}
\label{sec:rwork}

In this section, we briefly review related work on atomic multicast, geo-distributed systems, and high-performance recovery. 

\paragraph{Atomic multicast} 
Atomic multicast has been extensively studied in the literature.
In~\cite{BJ87b}, a protocol is proposed for failure-free scenarios. 
To decide on the final timestamp of a message, each process in the set of message addressees locally chooses a timestamp, exchanges its chosen timestamps, deterministically agrees on one of them, and delivers messages according to the message's final timestamp. 
Several works have extended this algorithm to tolerate failures~\cite{fritzke1998amcast, GS01b, rodrigues1998scalatom, schiper2008inherent}, where the main idea is to replace failure-prone processes by fault-tolerant disjoint groups of processes, each group implementing the algorithm by means of state-machine replication. 

Spread~\cite{ADM+04} implements a highly configurable group communication system, which supports the abstraction of process groups. 
Spread orders messages by the means of interconnected daemons that handle the communication in the system. 
Processes connect to a daemon to multicast and deliver messages.
%Similarly to Mencius~\cite{Mencius}, coordinators in Multi-Ring Paxos account for load imbalances by proposing skip messages in consensus instances. 
%Differently from Mencius, which is an atomic broadcast protocol, Multi-Ring Paxos implements atomic multicast by means of the abstraction of groups.
While the group abstraction is similar to the Totem Multi-Ring protocol~\cite{agarwal1998totem}, Totem uses timestamps to achieve global total 
order.
Multi-Ring Paxos's deterministic merge strategy is similar to the work proposed in \cite{aguilera2000detmerge}, which totally orders message streams in a widely distributed publish-subscribe system. 

% Atomic broadcast is not the only solution to totally order requests in a 
% distributed environment. 
% Distributed logging is an alternative approach, where appending a log entry 
% corresponds to executing a consensus instance in an atomic broadcast protocol. 
% CORFU~\cite{malkhi2012paxos} implements a distributed log with a cluster of 
% network-connected flash devices, where the log entries are partitioned among the 
% flash units. Each log entry is then made fault-tolerant using chain replication 
% and a set of flash devices. 
% Tango~\cite{balakrishnan2013tango}, builds on CORFU to implement partitioned 
% services, where a collection of log entries is allocated to each partition. 
% The replicas at each partition only execute the subset of the log entries 
% corresponding to their partitions, and skip the rest. Globally ordering the 
% entire set of log entries simplifies ensuring consistency with cross partition 
% queries. However, the number of partitions a service can be divided into is 
% limited by the log's capacity at handling the appends. 

\paragraph{Geo-replication}
There are different approaches to handling the high latency inherent of globally distributed systems. 
Some systems choose to weaken consistency guarantees (e.g., Dynamo~\cite{DeCandia07}), while others cope with wide-area round trip times.
Mencius~\cite{Mencius} and EPaxos~\cite{moraru2012egalitarian} are latency optimized. 
Both protocols implement atomic broadcast and therefore do not scale. 
P-store~\cite{schiper2010p} relies on atomic multicast.
In order to scale, it partitions the service state and strives to order requests that depend on each other, imposing a partial order on requests.
Sinfonia~\cite{aguilera2007sinfonia} and S-DUR~\cite{SPJ12} build a partial order by using a two-phase commit-like protocol to guarantee that 
requests spanning common partitions are processed in the same order at each partition.
Spanner~\cite{CDE12} orders requests within partitions using Paxos and across partitions using a protocol that computes a request's final timestamp from temporary timestamps proposed by the involved partitions.

\paragraph{Recovery} 
Recovery protocols often negatively affect a system's performance. 
Several optimizations can be applied to checkpointing and state transferring to minimize the overhead of recovery as we discuss next. 

% \emph{Optimized logging.} A common approach to efficient logging is to log 
% requests in 
% batches~\cite{bessani2013efficiency,castro1999practical,clement2009making,
% batching97,singh2009zeno}.
% Since stable storage devices are often block-based it is more efficient to write 
% a batch of requests into one block rather than to write multiple requests on 
% many different blocks. Another optimization is to parallelize the logging of 
% batches~\cite{bessani2013efficiency}. Parallel logging benefits most the 
% applications in which the time for processing a batch of requests is higher than 
% the time required for logging a batch. The overhead of logging can be further 
% reduced by using solid-state disks (SSD) or raw flash devices instead of 
% magnetic disks~\cite{rao2011using}. Similarly, in our dLog service we support 
% both harddisks and SSDs, and synchronous and asynchronous disk writes to enable 
% batched flushes to the disk. 

%\emph{Optimized checkpointing.} 
Some approaches have proposed to create checkpoints during the normal operation of a system, at the cost of halting normal command execution~\cite{lamport1998part,castro1999practical,rao2011using,singh2009zeno}. 
If all replicas stop simultaneously, the system becomes unavailable to clients and reduces performance. 
In~\cite{bessani2013efficiency} processes schedule checkpoints at different intervals, and therefore, the system is always operational.
As the operation of a quorum of processes is sufficient for their system to make progress, a minority of processes can be involved in a checkpointing while the other processes continue to operate. 
Another optimization is to use a \emph{helper} process to take checkpoints asynchronously~\cite{clement2009upright}. 
In this scheme, two threads, primary and the helper, execute concurrently. 
While the primary processes requests, the helper takes checkpoints periodically.

%\emph{Optimized state transfer.} 
State transfer has its own performance issues. 
During state transfer, the source process is involved both in the execution of commands and the transmission of the state to the recovering process, which may hamper performance. 
To address this problem, state transfer can be delayed to a moment when the demand on the system is low enough that both the execution of new requests and the transfer of the state can be handled~\cite{hunt2010zookeeper}. 
Another optimization is to reduce the amount of transferred information. 
Representing the state through efficient data structures~\cite{castro1999practical}, using incremental checkpoints~\cite{castro:2003base,clement2009upright}, or compressing the state are among these techniques. 
In~\cite{bessani2013efficiency}, the authors propose a collaborative state transfer protocol to evenly distribute the transfer load across replicas. 
RAMCloud~\cite{ousterhout2010case} is an in-memory storage system, where the data is also backed with persistent storage, such that the performance is not affected by the disk storage. 
To recover the data fast RAMCloud relies on the collective force of thousands of servers. 
%\hl{A similar speed-up can be achieved in our recovery procedure by using a distributed file system for storing and retrieving checkpoints.
%Different to our recovery mechanism, 
Differently from RAMCloud, our goal is to minimize application throughput disruption due to recovery, at the cost of a slower recovery procedure.
  
% Fast Databases with Fast Durability and Recovery Through Multicore Parallelism 

%% file: conclusion.tex
%!TEX root =  MRP-experience-report.tex
\section{Lessons learned}
\label{sec:lessons}

In this section, we present some lessons we learned during the implementation and evaluation of Multi-Ring Paxos. 
We first share our experiences with high throughput, low latency Java applications, and then consider issues more specific to Multi-Ring Paxos.

\subsection{High throughput and low latency in Java}

Implementing Multi-Ring Paxos in Java comes with benefits and challenges.
On the one hand, we can argue that Multi-Ring Paxos's code is easy to understand, maintain and extend.
We have evidence of this observation from users of Multi-Ring Paxos in our research group.
On the other hand, achieving high throughput in top-performing environments (e.g., 10 Gbps networks) and predictable low latency despite garbage collection is challenging.
%But when it comes to get out the best performance, like high network throughput of
%10 Gbit/s on a single Java VM, combined with predictable low latency garbage collection,
%things are getting complicated too. 

\myparagraph{Serialization.}
In the early versions of the URingPaxos library, serialization was a major performance problem.
To avoid getting tied to a specific serialization library, we decided to keep our own internal objects and then translate to whatever object was required by the serialization library.
This decision turned out to be problematic as, at high throughput, allocating all these extra objects caused a lot of garbage collection overhead.
We finally settled on implementing our own serialization using Java's byte buffers and avoiding the creation of extraneous objects.
%One major problem in early developments of the URingPaxos library was serialization. 
%\hl{To be serialization framework independent, we decided to keep internally our own object structure. }
%This decision requires that every network packet be de-serialized through the framework first and then copied to our object structure. 
% On high network traffic, the doubled object creation rate caused major garbage collection overhead. 
% In our current version, we implemented our own internal serialization based on byte buffers and create only one object per received message.

\myparagraph{In-memory storage.}
Acceptors can be configured to use on-disk or in-memory storage.
Using the in-memory storage (to avoid getting constrained by the performance of the disks), the acceptors have to keep enough data in memory to be able to handle the retransmission of recent messages (some seconds).
In a 10 Gbit network, that adds up to GBytes of memory that are constantly being replaced.
To avoid heavy garbage collection, we were keeping this data using a small library written in C, called through the Java Native Interface (JNI). It worked well but required a native, machine-dependent, library.
Our current implementation achieves the same result by pre-allocating a large array of byte buffers.
%\myparagraph{Buffer management.}
%While the incoming network traffic together with the serialization method influences the object creation rate, long-living objects, like the ones in the Paxos acceptor storage, cause different problems. 
%To evaluate the protocol performance and not benchmarking the underlying hard disks, the acceptors can use a in-memory storage back-end. 

% The in-memory storage, which does not constrain performance by the capacity of hard disks, keeps data enough to handle retransmissions of messages received in the last 3 to 5 seconds. 
% In a 10 Gbit network environment, this ends up in a Java memory heap of 3--5 GBytes, which is replaced frequently.
% To avoid heavy garbage collection work, the initial implementation used a procedure in C through the Java Native Interface (JNI) to allocate memory outside the Java heap. 
% This worked well but required a machine-dependent library. 
% The current implementation achieves the same results by pre-allocating an array of byte buffers. 
% Replacing the content of the buffers never triggers a mayor garbage collection effort.

\myparagraph{Garbage collection and heap size.}
While garbage collection does not significantly impact average throughput,
%For all experiments, we use Java default settings. 
its effect is clearly visible on latency measurements.
During our experiments, we observed that using smaller heap sizes resulted in smaller and more frequent GC pauses, leading to worse latency in average, but improving its standard deviation (i.e. short latency tail).
On the other hand, larger heap sizes caused less frequent but longer GC pauses, resulting in better latency in average but larger standard deviation (i.e. long latency tail).
This phenomenon can be observed in the garbage collection times CDF in Figures~\ref{fig:10g-mrp} and~\ref{fig:10g-200-mrp}.
The recovery experiment (Figure~\ref{fig:recovery}) also corroborates this idea: the large heap sizes we used incurred in GC pauses of up to a few seconds.

% Small heaps trigger the GC more often, which makes the average latency worse but improves the standard deviation. 
% Large heaps result in more infrequent GC runs and improve average latency, but result in a ``long latency tail''. 
% The phenomenon can be observed in the CDF of the garbage collection runtime in
% Figures~\ref{fig:10g-mrp} and~\ref{fig:10g-200-mrp}. 
% Further, in the recovery experiment, which used very large Java heaps, most of the visible impact is caused by GC (Figure~\ref{fig:recovery}).

\subsection{Protocol considerations}

We now consider aspects specifically related to Multi-Ring Paxos: its ring topology and recovery.

\myparagraph{Ring topology.}
While a ring topology helps achieving performance near nominal network capacity, it has the effect of propagating delays in a process to its successors. 
%All processes are communicating strictly in ring order. 
A pause in a single process (e.g. due to garbage collection or disk buffer flush) can cause the whole protocol to stop due to the serial propagation of messages in the ring.
Whenever possible, processes should first forward messages to its successor before doing any local processing.
%To improve latency, whenever possible incoming messages are first forwarded to the ring successor and then delivered locally.
%One single pause at a process (e.g., due to garbage collection or buffer flush) causes the whole serial ring execution to block.
%Therefore, sequential garbage collection or buffer flush events can cause severe 
%timeouts.
%But this approach is not always possible. 
%We are currently considering the use of deterministic garbage collection, a feature available in some real-time Java virtual machines.

\myparagraph{Recovery tradeoffs.}
Recovery involves many tradeoffs, which make the configuration of the protocol under high load difficult.
While in the original procedure recovery boils down to installing the most recent checkpoint and fetching missing commands from acceptors, in the
new procedure it involves caching new commands and waiting for a checkpoint that contains commands that precede the cached ones.
The new method does not place acceptors under additional stress (to recover missing commands) but it increases memory usage and management at the replicas.
Increased memory activity translates into new sources of overhead (e.g., garbage collection), which hurt performance.

The time needed for a checkpoint to be written to or read from stable storage introduces yet another tradeoff.
In order to reduce the number of commands cached by a recovering replica, the time for an operational replica to store a checkpointed state and for the recovering replica to fetch the stored checkpoint should be short.
Moreover, checkpoints must be created often.
Creating a checkpoint, however, introduces overheads during normal execution (although this is minimized by copy-on-write optimizations).

\section{Conclusions}
\label{sec:conclusion}

Internet-scale services rely on geographical distribution, data partitioning and replication to provide scalable performance and high availability.
Building such systems poses many challenges, one of them being the need to carefully order requests among replicas and partitions.
One way to cope with this problem is to use group communication primitives that encapsulate order requirements.
Multi-Ring Paxos is a protocol in this category.
Previous studies have considered Multi-Ring Paxos's performance in common environments.
In this paper, we consider the protocol in more extreme conditions. 
While these conditions can be considered exceptional, many systems already need to face them.
Therefore, understanding how Multi-Ring Paxos performs in such cases is important.
%We also reported on two novel optimizations that we introduced to boost Multi-Ring Paxos performance. !ENUMERATE?!

%We focused our performance study on three environments: a local environment (e.g., datacenter) with a 1 Gbps and a 10 Gbps network, and a global environment, where nodes are distributed across five geographical regions. 
%Multi-Ring Paxos showed good behavior under all evaluated conditions. 
%We could saturate a 10-Gbit network with ordered commands (of 32 Kbytes) and recover a replica with the system under peak load.
%Recovering under full load is a difficult problem and we had to carefully tune multiple parameters to achieve a non-disruptive procedure.
%The various tradeoffs involved in recovery in particular and Multi-Ring Paxos in general illustrate the obstacles that a high performance system must surmount to deliver on its promises.

%Recovering under full load is a difficult problem. 
%Multiple parameters must be carefully tuned to achieve non-disruptive recovery. 
%The various tradeoffs involving recovery in particular and Multi-Ring Paxos in general illustrate the obstacles that a high performance environment must surmount to deliver its promises.

%% file: MRP-experience-report.bbl
\begin{thebibliography}{10}

\bibitem{FLP85}
M.~J. Fischer, N.~A. Lynch, and M.~S. Paterson, ``Impossibility of distributed
  consensus with one faulty processor,'' {\em Journal of the ACM}, vol.~32,
  no.~2, pp.~374--382, 1985.

\bibitem{GILY2002}
S.~Gilbert and N.~Lynch, ``Brewer's conjecture and the feasibility of
  consistent, available, partition-tolerant web services,'' {\em SIGACT News},
  vol.~33, pp.~51--59, June 2002.

\bibitem{BJ87b}
K.~P. Birman and T.~A. Joseph, ``Reliable communication in the presence of
  failures,'' {\em ACM Transactions on Computer Systems (TOCS)}, vol.~5,
  pp.~47--76, Feb. 1987.

\bibitem{HT93}
V.~Hadzilacos and S.~Toueg, ``Fault-tolerant broadcasts and related problems,''
  in {\em Distributed Systems}, ch.~5, Addison-Wesley, 2nd~ed., 1993.

\bibitem{DUS04}
X.~D{\'e}fago, A.~Schiper, and P.~Urb{\'a}n, ``Total order broadcast and
  multicast algorithms: Taxonomy and survey,'' {\em ACM Computing Surveys,},
  vol.~36, pp.~372--421, Dec. 2004.

\bibitem{MPP2012}
P.~J. Marandi, M.~Primi, and F.~Pedone, ``Multi-ring paxos,'' in {\em DSN},
  2012.

\bibitem{BMP14}
S.~Benz, P.~J. Marandi, F.~Pedone, and B.~Garbinato, ``Building global and
  scalable systems with atomic multicast,'' in {\em Middleware}, 2014.

\bibitem{lamport1998part}
L.~Lamport, ``The part-time parliament,'' {\em ACM (TOCS)}, 1998.

\bibitem{RP2010}
P.~J. Marandi, M.~Primi, N.~Schiper, and F.~Pedone, ``Ring paxos: A
  high-throughput atomic broadcast protocol,'' in {\em DSN}, 2010.

\bibitem{Lam01}
L.~Lamport, ``Paxos made simple,'' {\em SIGACTN: SIGACT News (ACM Special
  Interest Group on Automata and Computability Theory)}, vol.~32, 2001.

\bibitem{DLS88}
C.~Dwork, N.~Lynch, and L.~Stockmeyer, ``Consensus in the presence of partial
  synchrony,'' {\em Journal of the ACM}, vol.~35, no.~2, pp.~288--323, 1988.

\bibitem{bessani2013efficiency}
A.~Bessani, M.~Santos, J.~Felix, N.~Neves, and M.~Correia, ``On the efficiency
  of durable state machine replication,'' in {\em ATC}, 2013.

\bibitem{hunt2010zookeeper}
P.~Hunt, M.~Konar, F.~P. Junqueira, and B.~Reed, ``Zookeeper: wait-free
  coordination for internet-scale systems,'' in {\em ATC}, 2010.

\bibitem{cooper2010benchmarking}
B.~F. Cooper, A.~Silberstein, E.~Tam, R.~Ramakrishnan, and R.~Sears,
  ``Benchmarking cloud serving systems with ycsb,'' in {\em SoCC}, 2010.

\bibitem{Jai91}
R.~Jain, {\em The art of computer systems performance analysis : techniques for
  experimental design, measurement, simulation, and modeling}.
\newblock New York: John Wiley and Sons, Inc., 1991.

\bibitem{fritzke1998amcast}
J.~Fritzke, U., P.~Ingels, A.~Mostefaoui, and M.~Raynal, ``Fault-tolerant total
  order multicast to asynchronous groups,'' in {\em SRDS}, 1998.

\bibitem{GS01b}
R.~Guerraoui and A.~Schiper, ``Genuine atomic multicast in asynchronous
  distributed systems,'' {\em Theor. Comput. Sci.}, vol.~254, no.~1-2,
  pp.~297--316, 2001.

\bibitem{rodrigues1998scalatom}
L.~Rodrigues, R.~Guerraoui, and A.~Schiper, ``Scalable atomic multicast,'' in
  {\em ICCCN}, 1998.

\bibitem{schiper2008inherent}
N.~Schiper and F.~Pedone, ``On the inherent cost of atomic broadcast and
  multicast in wide area networks,'' in {\em ICDCN}, 2008.

\bibitem{ADM+04}
Y.~Amir, C.~Danilov, M.~Miskin-Amir, J.~Schultz, and J.~Stanton, ``The {S}pread
  toolkit: Architecture and performance,'' tech. rep., Johns Hopkins
  University, 2004.
\newblock CNDS-2004-1.

\bibitem{agarwal1998totem}
D.~A. Agarwal, L.~E. Moser, P.~M. Melliar-Smith, and R.~K. Budhia, ``The totem
  multiple-ring ordering and topology maintenance protocol,'' {\em ACM}, May
  1998.

\bibitem{aguilera2000detmerge}
M.~K. Aguilera and R.~E. Strom, ``Efficient atomic broadcast using
  deterministic merge,'' in {\em PODC}, 2000.

\bibitem{DeCandia07}
G.~DeCandia, D.~Hastorun, M.~Jampani, G.~Kakulapati, A.~Lakshman, A.~Pilchin,
  S.~Sivasubramanian, P.~Vosshall, and W.~Vogels, ``Dynamo: {A}mazon's highly
  available key-value store,'' in {\em SOSP}, 2007.

\bibitem{Mencius}
Y.~Mao, F.~P. Junqueira, and K.~Marzullo, ``Mencius: building efficient
  replicated state machines for wans,'' in {\em OSDI}, 2008.

\bibitem{moraru2012egalitarian}
I.~Moraru, D.~G. Andersen, and M.~Kaminsky, ``Egalitarian paxos,'' in {\em
  SOSP}, 2012.

\bibitem{schiper2010p}
N.~Schiper, P.~Sutra, and F.~Pedone, ``P-store: Genuine partial replication in
  wide area networks,'' in {\em SRDS}, 2010.

\bibitem{aguilera2007sinfonia}
M.~K. Aguilera, A.~Merchant, M.~Shah, A.~Veitch, and C.~Karamanolis,
  ``Sinfonia: a new paradigm for building scalable distributed systems,'' in
  {\em SOSP}, 2007.

\bibitem{SPJ12}
D.~Sciascia, F.~Pedone, and F.~Junqueira, ``Scalable deferred update
  replication,'' in {\em DSN}, 2012.

\bibitem{CDE12}
J.~C. Corbett, J.~Dean, and M.~E. et~al, ``Spanner: Google's globally
  distributed database,'' in {\em OSDI}, 2012.

\bibitem{castro1999practical}
M.~Castro and B.~Liskov, ``Practical byzantine fault tolerance,'' in {\em
  OSDI}, 1999.

\bibitem{rao2011using}
J.~Rao, E.~J. Shekita, and S.~Tata, ``Using paxos to build a scalable,
  consistent, and highly available datastore,'' {\em Proceedings of the VLDB
  Endowment}, vol.~4, no.~4, pp.~243--254, 2011.

\bibitem{singh2009zeno}
A.~Singh, P.~Fonseca, P.~Kuznetsov, R.~Rodrigues, P.~Maniatis, {\em et~al.},
  ``Zeno: Eventually consistent byzantine-fault tolerance.,'' in {\em NSDI},
  2009.

\bibitem{clement2009upright}
A.~Clement, M.~Kapritsos, S.~Lee, Y.~Wang, L.~Alvisi, M.~Dahlin, and T.~Riche,
  ``Upright cluster services,'' in {\em SOSP}, 2009.

\bibitem{castro:2003base}
M.~Castro, R.~Rodrigues, and B.~Liskov, ``Base: Using abstraction to improve
  fault tolerance,'' {\em ACM Transactions on Computer Systems (TOCS)},
  vol.~21, no.~3, pp.~236--269, 2003.

\bibitem{ousterhout2010case}
J.~Ousterhout, P.~Agrawal, D.~Erickson, C.~Kozyrakis, J.~Leverich,
  D.~Mazi{\`e}res, S.~Mitra, A.~Narayanan, G.~Parulkar, M.~Rosenblum, {\em
  et~al.}, ``The case for ramclouds: scalable high-performance storage entirely
  in dram,'' {\em ACM SIGOPS OSR}, vol.~43, no.~4, pp.~92--105, 2010.

\end{thebibliography}
